\documentclass[prl,twocolumn,showpacs]{revtex4}
\usepackage{bm}
\usepackage{graphicx}
\usepackage{amssymb}
\usepackage{amsmath}
\usepackage{ulem}
\usepackage{color}
\newcommand{\nix}[1]{}

\begin{document}
\title{{Helicity sensitive terahertz radiation detection by field effect transistors}}

\author
{ C.~Drexler,$^1$  N. Dyakonova,$^2$ P. Olbrich,$^1$ J.~Karch,$^{1}$ M.~Schafberger,$^1$ K.~Karpierz,$^3$ Yu.~Mityagin,$^4$ M.~B.~Lifshits,$^{2,5}$ 
F.~Teppe,$^2$ O.~Klimenko,$^2$ Y.~M. Meziani,$^6$ W.~Knap,$^2$ and S.~D.~Ganichev$^{1}$}

\affiliation{$^1$ Terahertz Center, University of Regensburg,93040 Regensburg, Germany}

\affiliation{$^2$ Laboratoire Charles Coulomb UMR 5221 CNRS-UM2,Universite Montpellier 2, France}

\affiliation{$^3$ Institute of Experimental Physics, University of Warsaw, Hoza 69, 00-681 Warsaw, Poland}

\affiliation{$^4$ Lebedev Physical Institute, 119991, Leninsky prosp., 53, Moscow, Russia}

\affiliation{$^5$ A.F. Ioffe Physico-Technical Institute, 194021 St. Petersburg, Russia}

\affiliation{$^6$ Departamento de Fisica Aplicada, Universidad de Salamanca, E-37008 Salamanca, Spain}

\begin{abstract}
Terahertz light helicity sensitive photoresponse in GaAs/AlGaAs high electron mobility transistors. The helicity dependent detection mechanism is interpreted as an interference of plasma oscillations in the channel of the field-effect-transistors (generalized Dyakonov-Shur model). The observed helicity dependent photoresponse is by several orders of magnitude higher than any earlier reported one. Also linear polarization sensitive photoresponse was registered by the same transistors.  The results provide the basis for a new sensitive, all-electric, room-temperature and fast (better than 1 ns ) characterisation of all polarization parameters (Stokes parameters) of terahertz radiation.
%
%
It paves the way towards terahertz ellipsometry and polarization sensitive imaging based on 
plasma effects in field-effect-transistors. 
%
%
\end{abstract}



\date{\today}

\maketitle

\section{Introduction}
Terahertz (THz) science and technology hold a great promise for progress in diverse scientific areas and have a wide application potential in environmental monitoring, security, biomedical imaging and material characterization, see e.g.~\cite{sakai05,book,miles07,woolard07,Dexheimer2007,Lee2008,Zhang09,Mittleman2010}. Most of the potential THz applications require sensitive, but robust room temperature THz detectors with fast response time. Recently, field-effect-transistors (FETs) as well as low dimensional structures made of different semiconductor materials have been demonstrated as promising detectors of THz radiation~\cite{DS96,knapAPL02,knapJAP02,Knap02,peralta02,knap04,shaner05,Tauk06,sakowicz08,knap08,kim08,veksler09,knap09,dyer09,Popov11,Schuster11}. The operation principle is based on the use of nonlinear properties of the two-dimensional (2D) plasma in the transistor channel~\cite{DS96}. Both resonant~\cite{knapAPL02,Knap02} and nonresonant~\cite{knapJAP02} regimes of THz detection have been studied. Plasmonic effects can serve for room temperature detection from tens of gigahertz up to terahertz, enabling the combination of individual detectors in a matrix. 
While plasma nonlinearities based compact THz receivers are in focus of current research, the dependence of the voltage response on the radiation's polarization state is not yet exhaustively studied. The problem of radiation coupling and angular response to the linearly polarized radiation has been recently addressed by several groups~\cite{sakowicz08,knap08,kim08,Sak2010}. It was found that the response is proportional to the squared cosine of the azimuth angle and is attributed to the radiation coupling to the transistor via an antenna formed mainly by the bonding wires
and metallization of contact pads~\cite{sakowicz08,Sak2010}. However, no studies on the detector's response to circularly polarized radiation have been carried out so far.

\begin{figure}[b]
\includegraphics[width=0.8\linewidth]{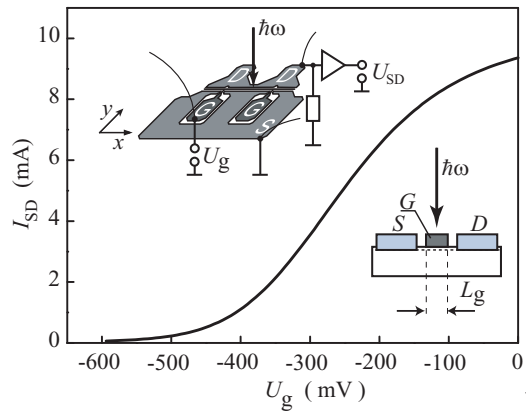}
\caption{Transfer
characteristic of GaAs/AlGaAs HEMT measured at $V_{\text{SD}}$ = 100~mV. Left inset shows experimental set-up and sketch of the device with contact pads S (source), D (drain) and G (gate). 
In optical experiments transistors were irradiated by elliptically, circularly or linearly polarized radiation at 
normal incidence. Right inset shows a sketch of the transistor cross section.}
\label{Fig:geometry}
\end{figure}

Here we report on the observation of the helicity dependent photoresponse 
of FETs. We demonstrate that for a certain design of GaAs/AlGaAs high electron mobility transistors (HEMTs) 
and silicon modulation-doped field effect transistors (Si-MODFETs) the photosignal has a substantial contribution proportional to the degree of circular polarization of the incident radiation. In particular, the photoresponse may change its sign when the light's helicity is switched from the right- to the left-handed circularly polarized state. The observed photoresponse is at least by three orders of magnitude higher than that known for the circular photogalvanic~\cite{Ganichev02,Ivchenko08,Ganichev03_1} and the spin-galvanic effect \cite{Ivchenko08,Ganichev03_2}. While the fact that terahertz plasmonic broadband detectors are sensitive to linear polarization is relatively well established~\cite{sakowicz08,knap08,kim08}, a photoresponse proportional to the degree of circular polarization has not been observed. We show that the generation of the photoresponse as well as its polarization dependence can be well described in the frame of the Dyakonov-Shur (DS) model of rectification by FETs~\cite{DS96}. Our results demonstrate that in HEMTs of particular design an access to the photon's helicity can be obtained through the interference of two \textit{ac} currents generated on opposite sides of the transistor channel. The observed helicity dependent photoresponse in FETs provides the basis for a sensitive all-electric characterization of THz radiation's polarization state and, therefore, can be used for the development of new methods of THz ellipsometry applying FET detectors. We emphasize that such room temperature FET detectors may have a very high sensitivity (5~kV/W) and low noise equivalent power below 10~pW$/\sqrt{\text{Hz}}$ \cite{Schuster11}. They can also be integrated on chip 
electronics and combined into matrices using standard III-V or Silicon CMOS technology~\cite{knap08,Lisauskas2009,Lisauskas2009_2,Knap2009_imaging}. 

\section{Experiment}

We studied commercially available GaAs/AlGaAs HEMTs with a gate length $L_\text{g}$ of 150~nm and a gate width $W_\text{g}$ of 280~$\mu$m. 
Figure~\ref{Fig:geometry} shows the design of the transistors, 
where the disposition of source (S), drain (D) and two gate (G) pads is sketched. The transistor's threshold voltage $U_{\text{th}}$ is obtained from the transfer characteristics, 
shown in  Fig.~\ref{Fig:geometry}, to approximately -440~mV. More detailed description of HEMT layers can be found in Ref.~[\onlinecite{transistor}]. We also investigated Si-MODFETs with $L_\text{g} \approx$ 150~nm and $W_\text{g} \approx 100~\mu$m.

\begin{figure}[b]
\includegraphics[width=1\linewidth]{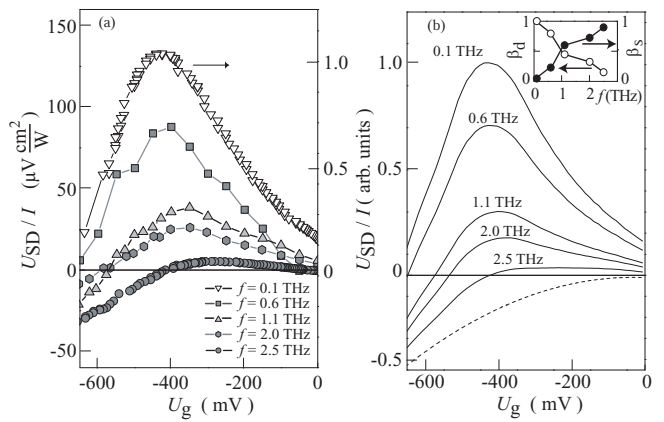}
\caption{
(a) Gate voltage dependence of the photovoltage induced in GaAs/AlGaAs HEMT by linearly polarized radiation. 
The data for frequencies $f=$  
1.1, 2.0, 2.5~THz  are 
multiplied  by factors 
5, 2, 10, respectively.
Note that the signals are read out after the inverting amplifier with a voltage amplification of 20~dB. 
(b) Gate voltage dependence calculated after Eq.~(\ref{Eq:six}) and assuming $C(\omega) = 1$. 
 Inset shows fitting parameters $\beta_\text{s}(f)$ and 
$\beta_\text{d}(f)$ as a function of the radiation frequency.  
}
\label{Fig:gatedep}
\end{figure}

The experiments have been performed applying a \textit{cw} methanol laser as well as pulsed NH$_{3}$, D$_2$O and CH$_{3}$F lasers~\cite{JETP1982,physicaB99} as sources of radiation in the frequency range of $f = 0.6 \div 2.5$~THz. The \textit{cw} laser operated at $f = 2.5$~THz (wavelength $\lambda = 118$~$\mu$m) with a power of about 20~mW~\cite{cw}. The NH$_{3}$, D$_{2}$O, CH$_{3}$F lasers provided 100~ns pulses at 2~THz ($\lambda = 148$~$\mu$m), 1.1~THz ($\lambda = 280$~$\mu$m), 0.8~THz ($\lambda = 385$~$\mu$m), and 0.6~THz ($\lambda = 496$~$\mu$m) with an output power of about 10~kW~\cite{Schneider2004,our6}. The incident radiation power was monitored by a photon drag reference detector~\cite{JTPL1985}. 
The laser beam was focused onto the sample by a parabolic mirror with a focal length of 65~mm. Typical laser spot diameters varied, depending on the wavelength, from 1 to 3~mm. The spatial laser beam distribution had an almost Gaussian profile, 
checked with a pyroelectric camera~\cite{Ziemann00}. In some experiments we used also a 95.5~GHz ($\lambda \approx 3.15$~mm) Gunn diode as a \textit{cw} source with a maximum output power of about 20~mW. No special antennas were used and the terahertz/microwave radiation was coupled to the device directly through bonding wires or metalization contact pads. 
The spatial distribution of the microwave radiation at 
the sample's position, and, in particular, the efficiency of the 
radiation coupling to the sample, by, e.g., 
the bonding wires and metalization of contact pads,
could not be measured. Thus, all microwaves data 
are given in arbitrary units.
Both linear and circular polarized light experiments were performed. To study the detector response upon variation of the radiation 
polarization state  $\lambda$/4 - and $\lambda$/2 - plates made of \textit{x}-cut crystalline quartz were used. 
In the pulsed THz experiments, the photoinduced voltage, $U_{\text{SD}}$, 
was fed through an inverting input of an amplifier with a voltage amplification of 20~dB and a bandwidth of 300~MHz (rise time of about 1~ns) to a digital broadband 1~GHz-oscilloscope. 
In the \textit{cw} experiments with the Gunn diode or gas laser sources, we used standard lock-in technique with a voltage amplification of 20~dB. The photoresponse was studied at different values of gate bias, $U_\text{g}$.
All experiments were performed at room temperature.

\begin{figure}[b]
\includegraphics[width=0.8\linewidth]{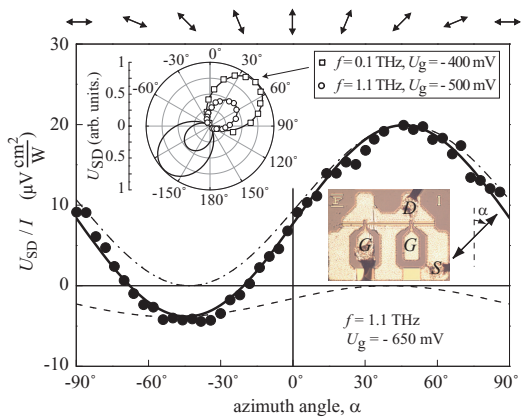}
\caption{Photovoltage as a function of the azimuth angle $\alpha$. Solid line shows the fit to Eq.~(\ref{Eq:one}). The dashed and dot-dashed lines display individual contributions proportional to $U_1 (U_\text{g}, f)$ and $U_2 (U_\text{g}, f)$, respectively. The left inset shows the polar plot of $U_{\text{SD}}(\alpha)$ measured for $f = 0.1$~THz, $U_\text{g} = -400$~mV (squares) and $f = 1.1$~THz, $U_\text{g} = -500$~mV (open circles) together with the corresponding fits to Eq.~(\ref{Eq:one}). 
Note that the maxima signals for $f=$~0.1~THz and 1.1~THz are normalized 
to 1 and 0.5, 
respectively. 
The right inset shows the picture of the transistor with bonding wires and defines the angle $\alpha$. 
On top the polarization direction corresponding to various azimuth angles are plotted. For zero angle $\alpha$ the radiation electric field vector is parallel to the $y$-axis.}
\label{Fig:alphadep}
\end{figure}

Results of experiments with linearly polarized light are summarized in Figs.~\ref{Fig:gatedep}(a) and~\ref{Fig:alphadep}.  
Figure~\ref{Fig:gatedep}(a) shows a gate voltage dependence of the HEMT's photoresponse measured for different radiation frequencies in the range of 0.1 to 2.5~THz.
For all used frequencies, the signal shows a maximum for a gate bias close to the threshold voltage, $U_{\text{th}} = - 440$~mV. With increasing frequency the magnitude of the maximum of the response decreases rapidly. The data show also that, at high frequencies, the signal switches its polarity at a certain gate bias. Note that the gate-bias of the inversion point diminishes with raising frequency. While a non-monotonic behavior of the signal, like observed for lower frequencies, is well known for FET detectors, the change of the response sign with  increasing radiation frequency has not been reported so far. 
Moreover, our experiments revealed a peculiar polarization behavior of the photosignal, in particular, at high radiation frequencies and large negative voltages. 

Sakowicz et al.~\cite{Sak2010} have shown that at low frequencies the radiation is coupled to the transistor mainly by bonding wires, whereas at higher frequencies ($>$ 100~GHz) the metalization of the contact pads plays the role of efficient antennas. Therefore, to interpret our  experiments we represent the photoinduced signal as a sum of two contributions: 
%
\begin{eqnarray}
U_{\text{SD}} = &U_1& (U_\text{g}, f) \cos^2 (\alpha + \theta_1) \label{Eq:one} \\
+&U_2& (U_\text{g}, f) \cos^2 (\alpha + \theta_2), \nonumber
\end{eqnarray}
with $U_1$ ($U_\text{g}$, $f$)  and $U_2$ ($U_\text{g}$, $f$) being of opposite sign. 
Herein, $\alpha$ is the azimuth angle defined in the inset of Fig.~\ref{Fig:alphadep}, and  $\theta_i$ are the phase angles, depending on the special alignment/geometry of bonding wires and metalization pads. Below we show that this assumption allows to reproduce the experimental traces, see Fig.~~\ref{Fig:alphadep}.
At the lowest radiation frequency of 100 GHz, the signal is well described by 
Eq.~(\ref{Eq:one}) with the phase angle $\theta_1 = 47^{\circ}$ and the second contribution close to zero. These data are shown in the top left inset in Fig.~\ref{Fig:alphadep}. 
Comparison of the photovoltage distribution
with the picture of the HEMT transistor (right inset in Fig.~\ref{Fig:alphadep})
reveals that the signal achieves a maximum value for the radiation polarization vector aligned roughly parallel to the line connecting the contact wires of the gate and drain. Disregarding the vanishingly small portion of the signal proportional to $U_2 (U_\text{g}, f)$, this behavior corresponds to the photoresponse of conventional FET detectors with the incoming radiation coupled to the transistor channel only by the bonding wires. Almost the same result has been obtained for higher frequencies and relatively small gate voltages, see e.g. the data for $f = 1.1$~THz and $U_\text{g} = -500$~mV shown in top left inset of Fig.~\ref{Fig:alphadep}. For higher frequencies and large bias voltages, however, we observed that the magnitudes of $\left|U_1 (U_\text{g}, f)\right|$ and $\left|U_2 (U_\text{g}, f)\right|$ become comparable. An example of such a behavior is shown in Fig.~\ref{Fig:alphadep} for $f =$ 1.1~THz and $U_\text{g} =$ -650 mV, together with the corresponding fits to Eq.~(\ref{Eq:one}). Note that the sign of 
$U_1 (U_\text{g}, f)$ and the value of the phase angle $\theta_1 = 47^{\circ}$ remain unchanged. The best fit is obtained for
negative value of $U_2 (U_\text{g}, f)$ and $\theta_2 = -50^{\circ}$,
i.e. the maximum of this contribution is obtained for the polarization vector oriented almost perpendicular to the line connecting the contact wires of the gate and drain. Thus, the contribution to the photosignal given by the voltage $U_2 (U_\text{g}, f)$ could not result from the 
antenna coupling to the connecting wires and can be only  attributed to the radiation coupling through the contact pads.

\begin{figure}[b]
\includegraphics[width=1\linewidth]{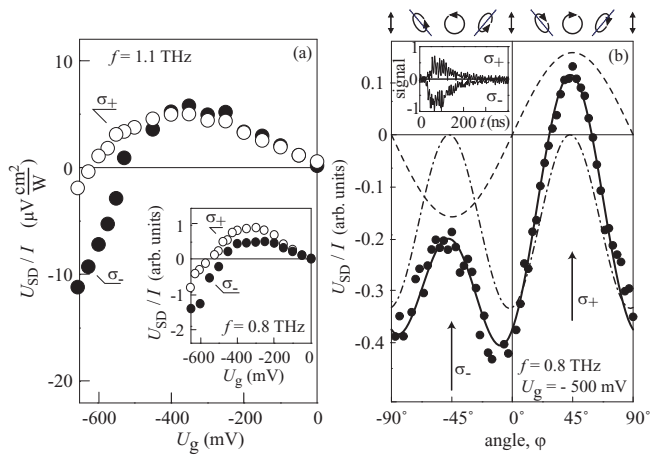}
\caption{(a) Photoresponse as a function of the gate voltage, measured for right- ($\sigma_+$) and left- ($\sigma_-$) circularly polarized radiation
with $f = 1.1$~THz.
Inset shows the data for $f =$ 0.8~THz. This data are normalized by the maximum signal in response to $\sigma_+$ radiation.
(b) Helicity dependence of the photovoltage measured for radiation with $f = 0.8$~THz. 
Solid line shows the fit to Eq.~(\ref{Eq:two}). The dashed and dot-dashed lines display individual contributions proportional to the parameters $U_\text{A}$ and $U_\text{B}$, respectively. 
 The ellipses on top illustrate the polarization states for various
$\varphi$.
The inset shows the photosignal pulse traces measured for $\sigma_+$ and $\sigma_-$ radiation.}
\label{Fig:phidep}
\end{figure}

Besides the complex polarization dependence upon rotation of the linear polarization plane, we observed that the photoresponse depends on the radiation's helicity. Fig´.~\ref{Fig:phidep}(a) shows $U_{\text{SD}}$ versus the gate voltage measured for right- ($\sigma_+$) and left- ($\sigma_-$) handed circularly polarized radiation with frequencies $f = 0.8$ and 1.1~THz. The data show that at relatively low bias voltages
, for which the dominating contribution of the signal proportional to $U_1 (U_\text{g}, f)$ is detected,
the photoresponse is insensitive to the radiation helicity. At higher negative bias voltages, however, the values and even the sign of the signal becomes different for $\sigma_+$ and $\sigma_-$ polarized radiation. By varying the radiation ellipticity we obtain a polarization behavior of the signal as shown Fig.~\ref{Fig:phidep}(b) which can be well reproduced by the following equation:
\begin{eqnarray}\label{Eq:two}
U_{\text{SD}} &=& U_\text{A} (U_\text{g}, f) \sin (2 \varphi) + \\
&& U_\text{B} (U_\text{g}, f) \cos^2 (2 \varphi + \theta) + U_\text{C} \:, \nonumber
\end{eqnarray} 
with $U_\text{A}$, $U_\text{B}$, $U_\text{C}$ and $\theta$ as the fitting parameters. This equation shows that the photoresponse is caused by a superposition of i) the signal proportional to the degree of circular polarization $P_{\text{circ}} = i(E_y E_x^* - E_x E_y^*)/\left| E \right|^2 = \sin{2 \varphi}$ (first term) and ii) signals determined by the degree of linear polarization of elliptically polarized light (second and third terms). Here, $\bm{E}$ is the radiation electric field. 
The inset in Fig.~\ref{Fig:phidep}(b) shows the typical photoresponse pulses for right- and left-handed circularly polarized radiation. A helicity dependent photoresponse was also observed for Si-MODFETs of similar design but the discussion of this data is out of scope of this paper.


\section{Theoretical background and discussion}

In the following, we show that the observed photoresponse as well as its polarisation dependence can be well described in the frame of the generalized model of THz detection by FETs~\cite{DS96}. The operation regime of the FET as a THz detector depends on the parameter $\omega \tau$, which determines whether plasma waves are excited ($\omega \tau > 1$) or not ($\omega \tau < 1$). In our experimental conditions the value of $\omega \tau$ is between 0.04 and 1 so that plasma oscillations are overdamped and the FET operates as a broadband detector. The channel length in the investigated transistors is much shorter than the wavelength, and the channel is mostly covered by the metallic gate. Hence, there is no direct interaction of the electromagnetic radiation with the electron gas in the channel. Therefore, the radiation is coupled to the channel through the source-gate and drain-gate contacts by effective antennas formed by the metallic contact pads and/or bonding wires. An \textit{ac} current induced by the radiation at the source (or at the drain) will leak to the gate at a distance on the order of the leakage length
%
\begin{equation}
l = \sqrt{\frac{2 \sigma}{ \omega C}} \:,
\label{Eq:three}
\end{equation}
where $\sigma$ is the conductivity of the channel and \textit{C} is the gate-to-channel capacitance per unit length. If the $L_\text{g}$ is much larger than the leakage length $l$, the photovoltage is generated in a region on the order of $l$ near the contact.

In the case when the radiation produces an \textit{ac} voltage with an amplitude $U_\text{s}$ between the source and gate contacts, the photoresponse $\Delta U$ is given by~\cite{D2010}:
\begin{eqnarray}
\Delta U &=& U_\text{s}^2 \cdot F(U_\text{g}),\\ 
F(U_\text{g}) &=& \frac{1}{4 ( U_\text{g} - U_\text{{th}})} \left[1- \frac{1}{2}\exp \left(-\frac{2 L_\text{g}}{l}\right)\right].
\label{Eq:four}
\end{eqnarray}
The prefactor in the expression for $F(U_\text{g})$ is valid for positive values of the gate voltage swing, $U_\text{g} - U_\text{th}$, which are not too close to zero.
 In the vicinity of the threshold and in the subthreshold region this prefactor was calculated in Ref.~\cite{DS96}. It depends on the \textit{dc} gate leakage current as well as on the transistor impedance. The dependence on $L_\text{g}/l$ is universal. Experimentally, the photoresponse as a function of $U_\text{g}$ has a broad maximum around the threshold. The amplitude of the input \textit{ac} voltage $U_\text{s}$ is related to the incoming radiation intensity $I$ and the antenna's sensitivity $\beta_\text{s} (\omega)$ as  $U_\text{s}^2 = \beta_\text{s} (\omega) \cdot I$. Thus, the photoresponse can be presented in the form~\cite{D2010}
\begin{equation}
	\Delta U = \beta_\text{s} (\omega)  \cdot
	F_\text{s}(U_\text{g}) \cdot I.
	\label{Eq:five}
\end{equation}
In order to interpret the experimental results, we assume that the radiation is fed to both, the source-gate and the gate-drain contacts. In other words, the coupling of the radiation to the transistor channel can be modeled by two effective antennas, one of them producing an \textit{ac} voltage between source and gate ($U_\text{s}$), and the other one between drain and gate ($U_\text{d}$). As long as $l$ is much smaller than $L_\text{g}$, there is no interference between the currents induced at opposite sides of the channel, and the corresponding contributions to the total photoresponse are independent. These contributions are obviously of opposite signs, resulting in a vanishing photoresponse for equal \textit{ac} amplitudes. The total photoresponse is given by
\begin{equation}
	\Delta U (U_\text{g}) = \left[\beta_\text{s} (\omega) 
	F_\text{s}(U_\text{g}) + \beta_d (\omega) 
	F_\text{d}(U_\text{g})\right] 
	\cdot C(\omega) 
	\cdot I,
	\label{Eq:six}
\end{equation}
where  $C(\omega)$ is the parameter describing the frequency dependence of the photoresponse, whereas
$\beta_\text{s}$ and $\beta_\text{d}$ are frequency dependent sensitivities of the source-gate and drain-gate effective antennas, 
respectively. The functions $F_\text{s}(U_\text{g})$ and $F_\text{d}(U_\text{g})$ describe the gate bias dependences of the photoresponse generated at the source and drain sides of the channel.

The role of antennas can be played by contact wires as well as metallic contact pads. Our results show that at $f$ = 95.5 GHz the incoming radiation is coupled to the transistor channel by contact wires. This fact is in agreement with Ref.~\cite{sakowicz08,Sak2010} where similar devices were studied. Indeed, we find that the photoresponse is optimal when the radiation polarization is aligned along the line connecting the gate and drain contact wires (see the right inset in Fig.~\ref{Fig:alphadep}). This demonstrates that at low frequencies the radiation is mostly coupled to the drain side of the channel. Hence, the upper curve in Fig.~\ref{Fig:gatedep}(a) can be attributed to the photoresponse generated at the drain side only, $F_\text{d}(U_\text{g})$, i.e. $\beta_\text{d}$ (0.1~THz) = 1 and $\beta_\text{s}$(0.1~THz) = 0. In contrast, at high frequencies, the radiation is expected to be coupled to the channel primarily through the contact metallic pads~\cite{Sak2010}. This enables coupling to the source side of the channel which, obviously, should lead to an opposite sign of the signal. Indeed, the photoresponse at 2.5~THz is negative in most of the gate voltage range. 
Figure~\ref{Fig:gatedep}(b) shows $\Delta U (U_\text{g})$ calculated for different fitting parameters $\beta_\text{s}(\omega)$ and $\beta_\text{d}(\omega)$. In order to determine the function $F_\text{d}(U_\text{g})$ we used the experimental dependence of the photoresponse on the gate voltage measured at $f \approx 0.1$~THz. Function $F_\text{s}(U_\text{g})$ is presented by the dashed line in Fig.~\ref{Fig:gatedep}(b).
In agreement with the above discussion it is assumed to be negative in the whole range of the gate voltage.

This choice of the functions $F_\text{s}(U_\text{g})$ and $F_\text{d}(U_\text{g})$ allows a good fit of photoresponse curves for all  
frequencies, see Fig.~\ref{Fig:gatedep}. The inset in Fig.~\ref{Fig:gatedep} demonstrates that the increase of the radiation frequency consistently decreases the efficiency of the gate-drain antenna, $\beta_\text{d}(\omega)$, and increases the coupling by the source-gate antenna, $\beta_\text{s}(\omega)$~\cite{footnote1}. Thus, at intermediate frequencies the photoresponse is a superposition of two signals generated at the drain and the source sides of the channel. This becomes also apparent in experiments with linear polarization of radiation. As seen in Fig.~\ref{Fig:alphadep}, at a frequency of 1.1~THz and $U_\text{g} = -650$~mV, the polarization dependence of the photoresponse is well fitted by the sum of dot-dashed and dashed curves~\cite{footnotex}. 

Now we turn to the observed helicity dependent photoresponse, and we show that it can just as well be understood using the described model of two effective antennas. So far we have assumed that the \textit{ac} currents generated on both sides of the channel do not interfere. This corresponds to the case when the leakage length $l$ is sufficiently small compared to $L_\text{g}/2$. 
If $l$ is of the order of $L_\text{g}/2$ there would be a region where the \textit{ac} currents generated at the source and drain coexist. Then the resulting \textit{ac} current will depend on their phase difference, $\xi$. Such a phase difference appears when source and drain are excited by mutually orthogonal components of circularly (or elliptically) polarized radiation. As we have seen above this is the case in our experimental setup, where source and drain effective antennas show maximal sensitivity for polarization directions differing by about 90$^\circ$. Consequently, the \textit{ac} currents generated at source and drain will  have phase shift $\xi$. It can be shown~\cite{LifhitsDyakonov} that, if $L_\text{g}$ is comparable to $l$, Eq.~(\ref{Eq:six}) should be modified as follows:
\begin{eqnarray}
&&	\Delta U = \left[\beta_\text{s} (\omega) 
F_\text{s}(U_\text{g}) + \beta_\text{d} (\omega) 
F_\text{d}(U_\text{g}) + \right.
\\  \nonumber
&& \left.
 \sqrt {\beta_\text{s} (\omega) 
 F_\text{s}(U_\text{g}) \beta_\text{d} (\omega) 
 F_\text{d}(U_\text{g})} \sin(\xi)\exp(-L_\text{g}/l)\right] \cdot I \:,
	\label{Eq:seven}
\end{eqnarray}
where the last interference term is sensitive to the radiation's helicity (sign of $\xi$). For linear polarization and $\xi = 0$, Eq.~(\ref{Eq:seven}) reduces to Eq.~(\ref{Eq:six}). The characteristic length is frequency dependent, 
see Eq.~(\ref{Eq:three}), and can be estimated to 76~nm (0.8~THz) and 89~nm (1.1~THz).
Thus, it is of the same order as the gate length, and the induced \textit{ac} currents at the drain and source sides can interfere.
As it follows from Fig. \ref{Fig:gatedep}, for $f = 1.1$~THz the contributions of source-gate and drain-gate antenna become equal at $U_\text{g} = -$565~mV, where $\beta_\text{s}(\omega) 
F_\text{s}(U_\text{g}) \approx \beta_\text{d}(\omega) 
F_\text{d}(U_\text{g})$. Thus, for this gate voltage only the interference term remains and the photoresponse should change its sign when the helicity of radiation is reversed. This is indeed what we observe, see Fig.~\ref{Fig:phidep}(a). 
The pulse traces obtained for right- and left- handed circularly polarized radiation demonstrate that the transistor allows the time resolved detection of a fine structure of the laser pulses with short spikes of the order of nanoseconds, see the inset in Fig.~\ref{Fig:phidep}(b). 
The response time of the transistor is determined by the time resolution of our setup, but it is 2~ns or less. The same response times were observed at $U_\text{g} = U_\text{th}$. The time constant is given by the cut-off frequency which is 10~GHz. The practically achievable time resolution, however, is \textit{RC}-limited by the design of the electric circuitry and by the bandwidth of cables and amplifiers.

\section{Summary}

To summarize, we demonstrate that for certain conditions high electron mobility transistors excited by terahertz laser radiation yield a photoresponse which depends on the radiation handedness, in particular may change its sign by reversing the circular polarization of radiation from right to left handed. The helicity dependent detection mechanism is interpreted in the frame of the generalized Dyakonov-Shur model, by taking into account the interference of the \textit{ac} currents in the transistor channel and specific antenna design. We show that, (i) optimization of the transistor design and (ii) the proper choice of the gate voltage, for which response to the linear polarization vanishes, should allow constructing a detector element with the responsivity just proportional to the radiation helicity, i.e. to the corresponding Stokes parameter. Combining such a detector element with two conventional FETs yielding a response to the linearly polarized radiation and, consequently, to the two remaining Stokes parameters should permit the complete characterization of radiation polarization in an all-electric manner. By the proper choice of the gate voltage, such room temperature detectors can be tuned to any radiation frequency in the THz range and are characterized by a nanosecond time resolution. Finally, we would like to emphasize that the observed transistor photoresponse is by several orders of magnitude higher than that one observed for the photogalvanic effect considered for the all-electric detection of the radiation's polarization state~\cite{Ganichev07,Danilov09}. Thus, the field effect transistor is a valuable candidate as a polarization detector. 

\section{Acknowledgements}

We thank  M.I. Dyakonov, V.V. Bel'kov,  and V.Yu. Kachorovskii for useful discussions. This work was supported by the DFG via Project SFB 689, Applications Center "Miniaturisierte Sensorik" (SappZ) of the Bavarian Government, the Linkage Grant of IB of BMBF at DLR. This work was supported by the CNRS through GDR-I project, by Region of Languedoc-Roussillon through the "Terahertz Platform" project and by the ANR via the "With" project. Y.~M.~M. acknowledges the financial help from the Ministry of Science and Innovation (project TEC2008-02281).


\begin{thebibliography}{99}
\bibitem{footnote1} It should be noted that here we compare only the shape of the photoresponse curves to establish the relative contributions of source and drain sides of the channel. The absolute value of the photoresponse depends also on the radiation power $P$, which, for the large variety of the radiation sources used in our work, could not be compared properly for different frequencies varying over almost two decades.
\bibitem{footnotex} Note that the first curve has a maximum when the polarization direction coincides with the line connecting the contact wires, i. e. it describes the photoresponse to the radiation coupled to drain side of the channel. The second one is of opposite sign and maximal when the polarization axis is turned by 97$^\circ$ with respect to the wires connecting line. This curve is attributed to the photoresponse caused by the radiation coupling to the source side of the channel.
\end{thebibliography}
\end{document}